\RequirePackage{lineno}

\documentclass[aps,prd,twocolumn,showpacs,superscriptaddress,groupedaddress]{revtex4}

\usepackage{graphicx}
\usepackage{dcolumn}
\usepackage{bm}

\begin{document}

\hspace{5.2in} \mbox{FERMILAB-PUB-19-542-E}

\title{Inclusive production of the {\boldmath $P_c$} resonances in {\boldmath $p \overline p $} collisions }

%
\affiliation{LAFEX, Centro Brasileiro de Pesquisas F\'{i}sicas, Rio de Janeiro, RJ 22290, Brazil}
\affiliation{Universidade do Estado do Rio de Janeiro, Rio de Janeiro, RJ 20550, Brazil}
\affiliation{Universidade Federal do ABC, Santo Andr\'e, SP 09210, Brazil}
\affiliation{University of Science and Technology of China, Hefei 230026, People's Republic of China}
\affiliation{Universidad de los Andes, Bogot\'a, 111711, Colombia}
\affiliation{Charles University, Faculty of Mathematics and Physics, Center for Particle Physics, 116 36 Prague 1, Czech Republic}
\affiliation{Czech Technical University in Prague, 116 36 Prague 6, Czech Republic}
\affiliation{Institute of Physics, Academy of Sciences of the Czech Republic, 182 21 Prague, Czech Republic}
\affiliation{Universidad San Francisco de Quito, Quito 170157, Ecuador}
\affiliation{LPC, Universit\'e Blaise Pascal, CNRS/IN2P3, Clermont, F-63178 Aubi\`ere Cedex, France}
\affiliation{LPSC, Universit\'e Joseph Fourier Grenoble 1, CNRS/IN2P3, Institut National Polytechnique de Grenoble, F-38026 Grenoble Cedex, France}
\affiliation{CPPM, Aix-Marseille Universit\'e, CNRS/IN2P3, F-13288 Marseille Cedex 09, France}
\affiliation{LAL, Univ. Paris-Sud, CNRS/IN2P3, Universit\'e Paris-Saclay, F-91898 Orsay Cedex, France}
\affiliation{LPNHE, Universit\'es Paris VI and VII, CNRS/IN2P3, F-75005 Paris, France}
\affiliation{IRFU, CEA, Universit\'e Paris-Saclay, F-91191 Gif-Sur-Yvette, France}
\affiliation{IPHC, Universit\'e de Strasbourg, CNRS/IN2P3, F-67037 Strasbourg, France}
\affiliation{IPNL, Universit\'e Lyon 1, CNRS/IN2P3, F-69622 Villeurbanne Cedex, France and Universit\'e de Lyon, F-69361 Lyon CEDEX 07, France}
\affiliation{III. Physikalisches Institut A, RWTH Aachen University, 52056 Aachen, Germany}
\affiliation{Physikalisches Institut, Universit\"at Freiburg, 79085 Freiburg, Germany}
\affiliation{II. Physikalisches Institut, Georg-August-Universit\"at G\"ottingen, 37073 G\"ottingen, Germany}
\affiliation{Institut f\"ur Physik, Universit\"at Mainz, 55099 Mainz, Germany}
\affiliation{Ludwig-Maximilians-Universit\"at M\"unchen, 80539 M\"unchen, Germany}
\affiliation{Panjab University, Chandigarh 160014, India}
\affiliation{Delhi University, Delhi-110 007, India}
\affiliation{Tata Institute of Fundamental Research, Mumbai-400 005, India}
\affiliation{University College Dublin, Dublin 4, Ireland}
\affiliation{Korea Detector Laboratory, Korea University, Seoul, 02841, Korea}
\affiliation{CINVESTAV, Mexico City 07360, Mexico}
\affiliation{Nikhef, Science Park, 1098 XG Amsterdam, the Netherlands}
\affiliation{Radboud University Nijmegen, 6525 AJ Nijmegen, the Netherlands}
\affiliation{Joint Institute for Nuclear Research, Dubna 141980, Russia}
\affiliation{Institute for Theoretical and Experimental Physics, Moscow 117259, Russia}
\affiliation{Moscow State University, Moscow 119991, Russia}
\affiliation{Institute for High Energy Physics, Protvino, Moscow region 142281, Russia}
\affiliation{Petersburg Nuclear Physics Institute, St. Petersburg 188300, Russia}
\affiliation{Instituci\'{o} Catalana de Recerca i Estudis Avan\c{c}ats (ICREA) and Institut de F\'{i}sica d'Altes Energies (IFAE), 08193 Bellaterra (Barcelona), Spain}
\affiliation{Uppsala University, 751 05 Uppsala, Sweden}
\affiliation{Taras Shevchenko National University of Kyiv, Kiev, 01601, Ukraine}
\affiliation{Lancaster University, Lancaster LA1 4YB, United Kingdom}
\affiliation{Imperial College London, London SW7 2AZ, United Kingdom}
\affiliation{The University of Manchester, Manchester M13 9PL, United Kingdom}
\affiliation{University of Arizona, Tucson, Arizona 85721, USA}
\affiliation{University of California Riverside, Riverside, California 92521, USA}
\affiliation{Florida State University, Tallahassee, Florida 32306, USA}
\affiliation{Fermi National Accelerator Laboratory, Batavia, Illinois 60510, USA}
\affiliation{University of Illinois at Chicago, Chicago, Illinois 60607, USA}
\affiliation{Northern Illinois University, DeKalb, Illinois 60115, USA}
\affiliation{Northwestern University, Evanston, Illinois 60208, USA}
\affiliation{Indiana University, Bloomington, Indiana 47405, USA}
\affiliation{Purdue University Calumet, Hammond, Indiana 46323, USA}
\affiliation{University of Notre Dame, Notre Dame, Indiana 46556, USA}
\affiliation{Iowa State University, Ames, Iowa 50011, USA}
\affiliation{University of Kansas, Lawrence, Kansas 66045, USA}
\affiliation{Louisiana Tech University, Ruston, Louisiana 71272, USA}
\affiliation{Northeastern University, Boston, Massachusetts 02115, USA}
\affiliation{University of Michigan, Ann Arbor, Michigan 48109, USA}
\affiliation{Michigan State University, East Lansing, Michigan 48824, USA}
\affiliation{University of Mississippi, University, Mississippi 38677, USA}
\affiliation{University of Nebraska, Lincoln, Nebraska 68588, USA}
\affiliation{Rutgers University, Piscataway, New Jersey 08855, USA}
\affiliation{Princeton University, Princeton, New Jersey 08544, USA}
\affiliation{State University of New York, Buffalo, New York 14260, USA}
\affiliation{University of Rochester, Rochester, New York 14627, USA}
\affiliation{State University of New York, Stony Brook, New York 11794, USA}
\affiliation{Brookhaven National Laboratory, Upton, New York 11973, USA}
\affiliation{Langston University, Langston, Oklahoma 73050, USA}
\affiliation{University of Oklahoma, Norman, Oklahoma 73019, USA}
\affiliation{Oklahoma State University, Stillwater, Oklahoma 74078, USA}
\affiliation{Oregon State University, Corvallis, Oregon 97331, USA}
\affiliation{Brown University, Providence, Rhode Island 02912, USA}
\affiliation{University of Texas, Arlington, Texas 76019, USA}
\affiliation{Southern Methodist University, Dallas, Texas 75275, USA}
\affiliation{Rice University, Houston, Texas 77005, USA}
\affiliation{University of Virginia, Charlottesville, Virginia 22904, USA}
\affiliation{University of Washington, Seattle, Washington 98195, USA}
\author{V.M.~Abazov} \affiliation{Joint Institute for Nuclear Research, Dubna 141980, Russia}
\author{B.~Abbott} \affiliation{University of Oklahoma, Norman, Oklahoma 73019, USA}
\author{B.S.~Acharya} \affiliation{Tata Institute of Fundamental Research, Mumbai-400 005, India}
\author{M.~Adams} \affiliation{University of Illinois at Chicago, Chicago, Illinois 60607, USA}
\author{T.~Adams} \affiliation{Florida State University, Tallahassee, Florida 32306, USA}
\author{J.P.~Agnew} \affiliation{The University of Manchester, Manchester M13 9PL, United Kingdom}
\author{G.D.~Alexeev} \affiliation{Joint Institute for Nuclear Research, Dubna 141980, Russia}
\author{G.~Alkhazov} \affiliation{Petersburg Nuclear Physics Institute, St. Petersburg 188300, Russia}
\author{A.~Alton$^{a}$} \affiliation{University of Michigan, Ann Arbor, Michigan 48109, USA}
\author{A.~Askew} \affiliation{Florida State University, Tallahassee, Florida 32306, USA}
% ...
%
%
\collaboration{The D0 Collaboration\footnote{with visitors from
$^{a}$Augustana College, Sioux Falls, SD 57197, USA,
$^{b}$The University of Liverpool, Liverpool L69 3BX, UK,
$^{c}$Deutshes Elektronen-Synchrotron (DESY), Notkestrasse 85, Germany,
$^{d}$CONACyT, M-03940 Mexico City, Mexico,
$^{e}$SLAC, Menlo Park, CA 94025, USA,
$^{f}$University College London, London WC1E 6BT, UK,
$^{g}$Centro de Investigacion en Computacion - IPN, CP 07738 Mexico City, Mexico,
$^{h}$Universidade Estadual Paulista, S\~ao Paulo, SP 01140, Brazil,
$^{i}$Karlsruher Institut f\"ur Technologie (KIT) - Steinbuch Centre for Computing (SCC),
D-76128 Karlsruhe, Germany,
$^{j}$Office of Science, U.S. Department of Energy, Washington, D.C. 20585, USA,
$^{l}$Kiev Institute for Nuclear Research (KINR), Kyiv 03680, Ukraine,
$^{m}$University of Maryland, College Park, MD 20742, USA,
$^{n}$European Orgnaization for Nuclear Research (CERN), CH-1211 Geneva, Switzerland,
$^{o}$Purdue University, West Lafayette, IN 47907, USA,
$^{p}$Institute of Physics, Belgrade, Belgrade, Serbia,
and
$^{q}$P.N. Lebedev Physical Institute of the Russian Academy of Sciences, 119991, Moscow, Russia.
$^{\ddag}$Deceased.
}} \noaffiliation
\vskip 0.25cm

\date{\today}
           
\begin{abstract}

We present a study of the inclusive production in  $p \overline p $ collisions 
of the  pentaquark states $P_c(4440)$ and $P_c(4457)$ with the decay to the
$J/\psi p$ final state previously observed by the LHCb experiment.
Using a sample of candidates originating from decays of $b$-flavored hadrons,
we find  an enhancement in the $J/\psi p$ invariant mass
distribution consistent with the sum of  $P_c(4440)$ and $P_c(4457)$.
The significance, with the mass and width  parameters set  to the LHCb measured values
and including the D0 systematic uncertainties and uncertainties in the LHCb input parameters
for the $P_c(4440)$ and $P_c(4457)$, is $3.2\sigma$.
The study of the semi-exclusive process $\Lambda_b \rightarrow J/\psi p X$
indicates the possibility that decays $\Lambda_b \rightarrow P_c  X$
other than those with  $X=K^-$ exist.
This is the first confirmatory evidence for these pentaquark states.
We measure the ratio $N_{\rm prompt}/N_{\rm nonprompt}=0.1 \pm 0.4$
and set an upper limit  of 0.9 at the  95\% credibility level.
 The ratio of the yield of the $P_c(4312)$ to the sum of $P_c(4440)$ and $P_c(4457)$,
measured to be 0.18$\pm$0.22, is consistent with the LHCb result. 
The study is based on  $10.4~\rm{fb^{-1}}$ of
data  collected by the  D0 experiment at  the Fermilab Tevatron collider. 
 
\end{abstract}

\pacs{14.40.Cx,13.25.Cv,12.39.Mk}

\maketitle

Since the discovery~\cite{belle2003} of the charmonium-like state  $\chi_{c1}(3872)$ 
(known also as  $X(3872)$) in 2003,  evidence
 has accumulated for mesons with a quark content
beyond the  color-singlet quark-antiquark combination.
Currently,  a dozen mesons require presence 
of a hidden charm $c\overline c$ pair in addition
to a light-quark  $q \bar q$ pair.
Several models have been proposed to describe the internal structure
of a multi-quark state:
a compact heavy quark-antiquark core surrounded by a light-quark cloud;
 a bound state of hypothetical compact  diquarks;
a deuteron-like hadronic molecule composed of two color-singlet heavy hadrons. For recent reviews
 see Refs. 
 ~\cite{osz,ghm,krs,lcc}.

Until recently,   there have been no undisputed
``pentaquark'' baryons.
That changed in  2015 with a discovery~\cite{lhcbpc}  by  the LHCb Collaboration
 of  two particles decaying
to $J/\psi p$.
The minimum quark content of such a state  is $c \overline c uud$ 
(charge conjugation is implied throughout this paper).
Recently, using an increased dataset, the LHCb Collaboration reported the discovery of three narrow
resonances~\cite{lhcb2019}  in the $J/\psi p$ invariant mass spectrum,  
$P_c(4312)$, $P_c(4440)$, and $P_c(4457)$ with the following
mass  and width parameters:
\vskip 1mm

 $M=4311.9 \pm 0.7 ^{+6.8}_{-0.6}$~MeV, $\Gamma=9.8 \pm 2.7^{+3.7}_{-4.5}$~MeV

 $M=4440.3 \pm 1.3 ^{+4.1}_{-4.7}$~MeV, $\Gamma=20.6 \pm 4.9^{+8.7}_{-10.1}$~MeV

 $M=4457.3 \pm 0.6 ^{+4.1}_{-1.7}$~MeV, $\Gamma=6.4 \pm 2.0^{+5.7}_{-1.9}$~MeV.

\vskip 1mm
\noindent These new results supersede those previously presented  in Ref.~\cite{lhcbpc}.

The $P_c$  states were found as resonances in the decay products of the $\Lambda_b^0$ baryon,
$\Lambda_b^0 \rightarrow J/\psi p K^-$, and also in  $\Lambda_b^0 \rightarrow J/\psi p \pi^-$~\cite{lhcbpcpi}.
They might also  be produced in other  $\Lambda_b^0$  
decay channels, such as $\Lambda_b^0 \rightarrow J/\psi p K^{*-}$
or any channel containing one or two pions in addition to 
$J/\psi p K^{-(0)}$, 
 in decays of other $b$ hadrons ($H_b$) 
or promptly in gluon-gluon  or quark-antiquark fusion.
In this Article  we present results of a search for the inclusive
production of the $P_c$ states
in  $p \overline p $ collisions.
Due to limited mass resolution and high background, this study is focused on
 a search for  a  signal consisting of a  sum of
 the  $P_c(4440)$ and $P_c(4457)$  resonances with the mass and width parameters
taken from  Ref.~\cite{lhcb2019}.

 Although the observation of the $P_c$ states in Refs.~\cite{lhcbpc,lhcb2019} was
 in the decay channel $\Lambda_b^0 \rightarrow  P_c^+K^-$ with
 $P_c \rightarrow J/\psi p$, our analysis was initially  based on the inclusive
 $J/\psi p$ sample so as to allow
 the contributions from other $\Lambda_b$ decays  and from $b$-quark
 meson states.
 We also discuss the study of the exclusive 
 $\Lambda_b^0 \rightarrow J/\psi p K^-$  channel and confirm that it yields 
lower significance than the inclusive channel.

The data sample,  corresponding to an integrated luminosity of 10.4~fb$^{-1}$,
was  collected with the D0 detector  in $p \overline p $ collisions at 1.96 TeV at the Fermilab Tevatron 
collider.

The D0 detector has a central tracking system consisting of a silicon
microstrip tracker  and a central fiber tracker, both located within a
1.9~T superconducting solenoidal magnet
 and a liquid argon calorimeter~\cite{d0det, layer0, calo}. A muon
system, covering $|\eta|<2$~\cite{eta}, consists of a layer of tracking
detectors and scintillation trigger counters in front of a central and two forward 
1.8~T iron toroidal
magnets, followed by two similar layers after the toroids~\cite{run2muon}.
Events used in this analysis are collected with both single-muon and dimuon triggers.
Single-muon triggers require a coincidence of signals in trigger elements inside and
 outside
the toroidal magnets.
 All dimuon triggers  require at least one muon to have track segments  after the toroid;
muons in the forward region are always required to penetrate the   toroid.
The minimum muon transverse momentum is 1.5~GeV.
No minimum $p_T$ requirement is applied to the muon pair,
but the effective threshold is approximately 4~GeV due to the requirement for muons
to penetrate the toroids,  and the average value for
accepted events is 10~GeV.

The event selection, detailed below,  follows that of Refs.~\cite{zc1,d0zc2}
that reported evidence for the presence of the decay process 
$Z_c(3900) \rightarrow J/\psi \pi$ in the final states
of $b$-hadron decays. 
Candidate events  are selected  by requiring  a pair of oppositely charged  muons
and  a charged particle   at a common vertex  with
 $\chi^2 < 10$ for 3 degrees of freedom.
Muons  must
   have transverse momentum $p_T > 1.5$~GeV.
At least one muon must traverse both inner and outer layers of the muon detector.
 Both muons must match tracks
in the central tracking system. The reconstructed invariant mass $M(\mu^+\mu^-)$
must be between 2.92 and 3.25~GeV,
consistent with the world average mass of the $J/\psi$~\cite{pdg}.

There are  several differences appropriate for this analysis.
The charged particle accompanying the $J/\psi$ candidate is required to have 
transverse momentum   $p_T>2$~GeV and  is assigned the proton mass.
The increased limit on the particle  $p_T$ is based upon the kinematic fact that in particle decays, 
a heavy particle (e.g. a proton) carries more momentum than a light particle (e.g. a pion)
According to simulations, the lower limit of 2~GeV enhances
the decays of $\Lambda_b^0$ over decays of other $b$ hadrons by a factor of
about two. It also suppresses background from pairing a $J/\psi$ produced in a $b$-hadron decay
with a random low-$p_T$ hadron from hadronization.

We set an upper limit on the $J/\psi p$ transverse momentum based upon the observation
 that the $p_T$ distribution of $\Lambda_b^0$'s is softer than that for $B$ mesons~\cite{ptlb},
 and on our expectation that a $P_c$ signal has a dominant contribution from $\Lambda_b^0$
  decays.
 Our unpublished study of the decay $\Lambda_b^0 \rightarrow  J/\psi \Lambda$ shows that 
the fraction of accepted $\Lambda_b^0$~s with $p_T > 13$ GeV is less than 10\%.
  On average we expect the other particle from $\Lambda_b^0$  decay ($K$, $K^*$ etc.)
 to contribute a $p_T$ of about 1 GeV, leading to our choice of 12 GeV for the $J/\psi  p$
 transverse momentum thus retaining an estimated 90\% of $\Lambda_b^0$'s.

 The invariant mass of the $J/\psi p$ candidate
is limited to the  range 4.2--4.6~GeV.

\begin{figure}[htb]
\includegraphics[scale=0.42]{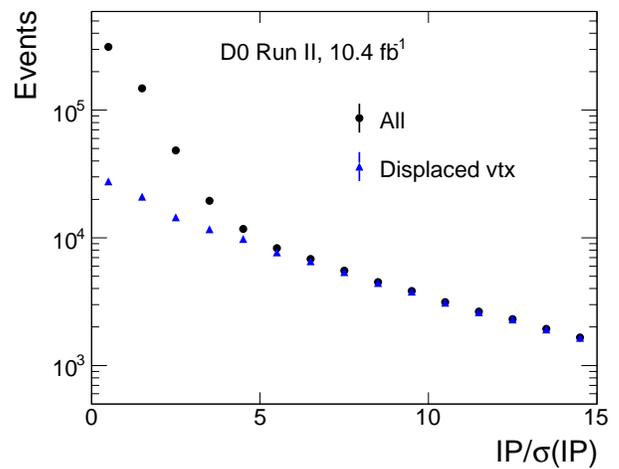}
\caption{\label{fig:ip}
 Impact parameter significance
of the proton candidate  before (black circles) and after (blue triangles) the requirement
of    the $J/\psi p$ vertex 
 separation 
from the primary vertex.
}
\end{figure}

Similarly to Refs.~\cite{zc1,d0zc2},
to select events where the $J/\psi p$ pair   originates from a $b$-hadron decay,
the  $J/\psi p$  vertex is required to be  displaced  in the transverse plane from
the $p\bar{p}$  interaction vertex  by at least 5$\sigma$.
The combination of the above requirements suppresses promptly produced $J/\psi$
 proton candidates.   Figure ~\ref{fig:ip} indicates that a further cut on proton
 impact parameter~\cite{ip}, as
 employed in Refs.~\cite{zc1,d0zc2} is not necessary.  Our analysis assumes no leakage
from prompt $J/\psi p$ events.
The resulting ``displaced vertex'' sample  contains 137,357 events.

In a search  for $P_c$ states coming from $b$-hadron decays, we study the  $M(J/\psi p)$
distribution of the ``displaced vertex'' events. 
We perform binned maximum likelihood fits 
assuming a  signal described below, convolved with a Gaussian resolution,
and a baseline background choice of a second-order Chebyshev polynomial.
The choice of a smooth background shape is justifed by the absence of narrow peaking background.
 Contributions from $P_c(4312)$ or
any peaking background  are neglected. However,  including those  states would not diminish the fitted signal.
Reflections from $J/\psi \pi$ resonances make contributions that are wider than the parent state
and can be incorporated in the polynomial background.
At around 4.45~GeV, the mass resolution is 12$\pm$2~MeV. 
As a test of the smooth background parametrization we have conducted
a likelihood scan using
data in  the mass range 4.2--4.4~GeV,  assuming several values of the resonance mass.
The largest deviation from zero among 13 points is $S=1.7 \sigma$,
consistent with a  background-only behavior.
The fit quality for null signal is $\chi^2/ndof=20/17$.
We conclude that the Chebyshev polynomial background model is adequate
for our data.

We treat  the signal near 4.45~GeV as an incoherent  sum  of the $P_c(4440)$ and $P_c(4457)$ Breit-Wigner
 resonances, with  the mass and width parameters equal  to the LHCb values.  
We allow 
the relative contribution
of the two yields,   
$f=N(4440)/(N(4440)+N(4457))$ to vary.
Our assumption of an incoherent sum of the $P_c$ states is based on the theoretical predictions
that  these  two states have different $J^P$ values.
They have been widely discussed as $\Sigma_c \bar D^*$ molecules or compact
diquark structures.
In the molecular picture, the $J^P$ of the $P_c(4440)$ and $P_c(4457)$
can be $[1/2^-,3/2^-]$ or $[3/2^-,1/2^-]$~\cite{meng}.
In the compact diquark model, the $J^P$ of
 $P_c(4440)$ and $P_c(4457)$ are  $[3/2^+,5/2^+]$~\cite{diquark}.

\begin{figure}[htbp]
  \includegraphics[width=0.99\columnwidth,height=7.2cm]{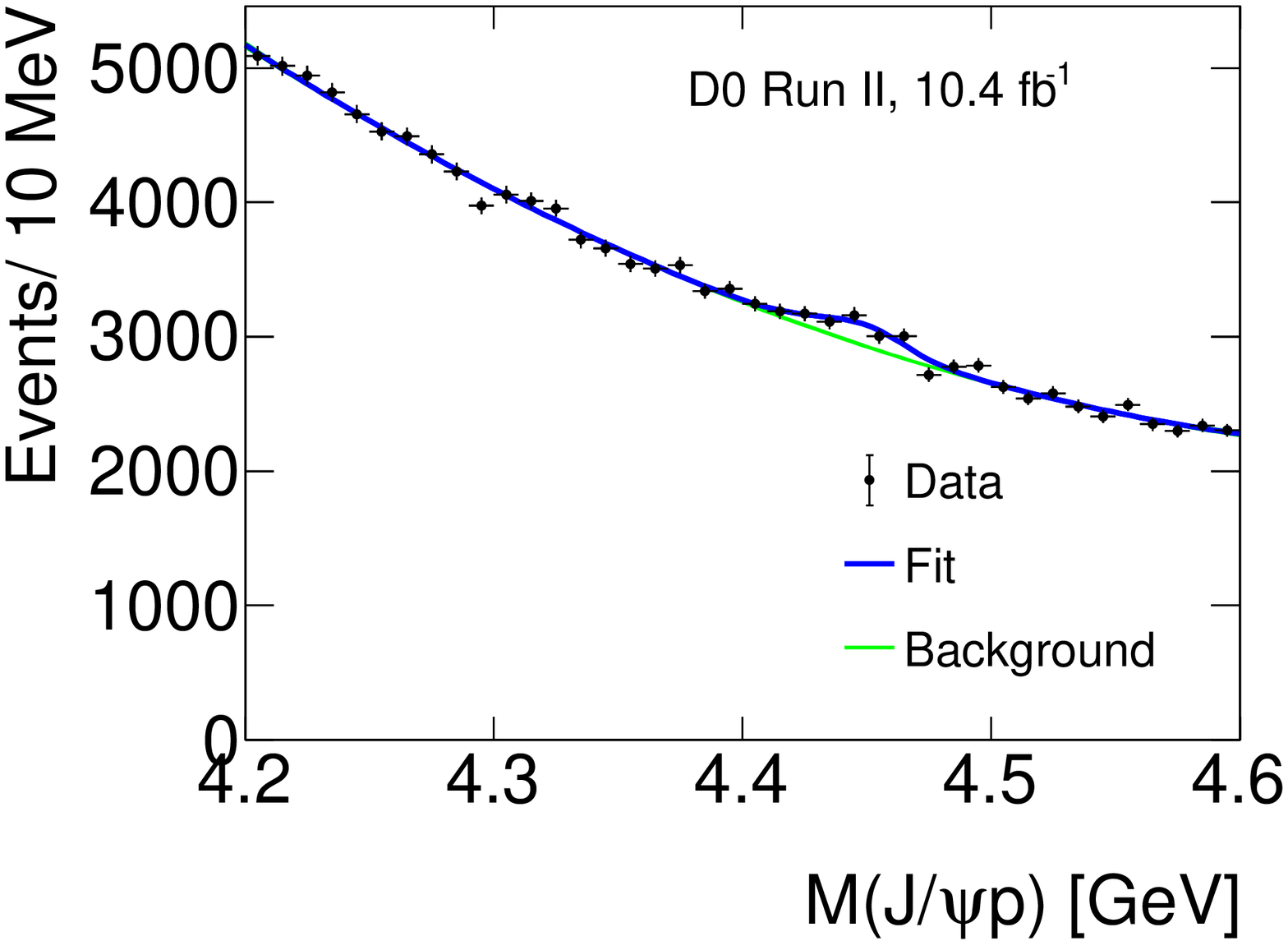}
\vskip-4.25cm\hskip-1.4cm\includegraphics[height=2.75cm]{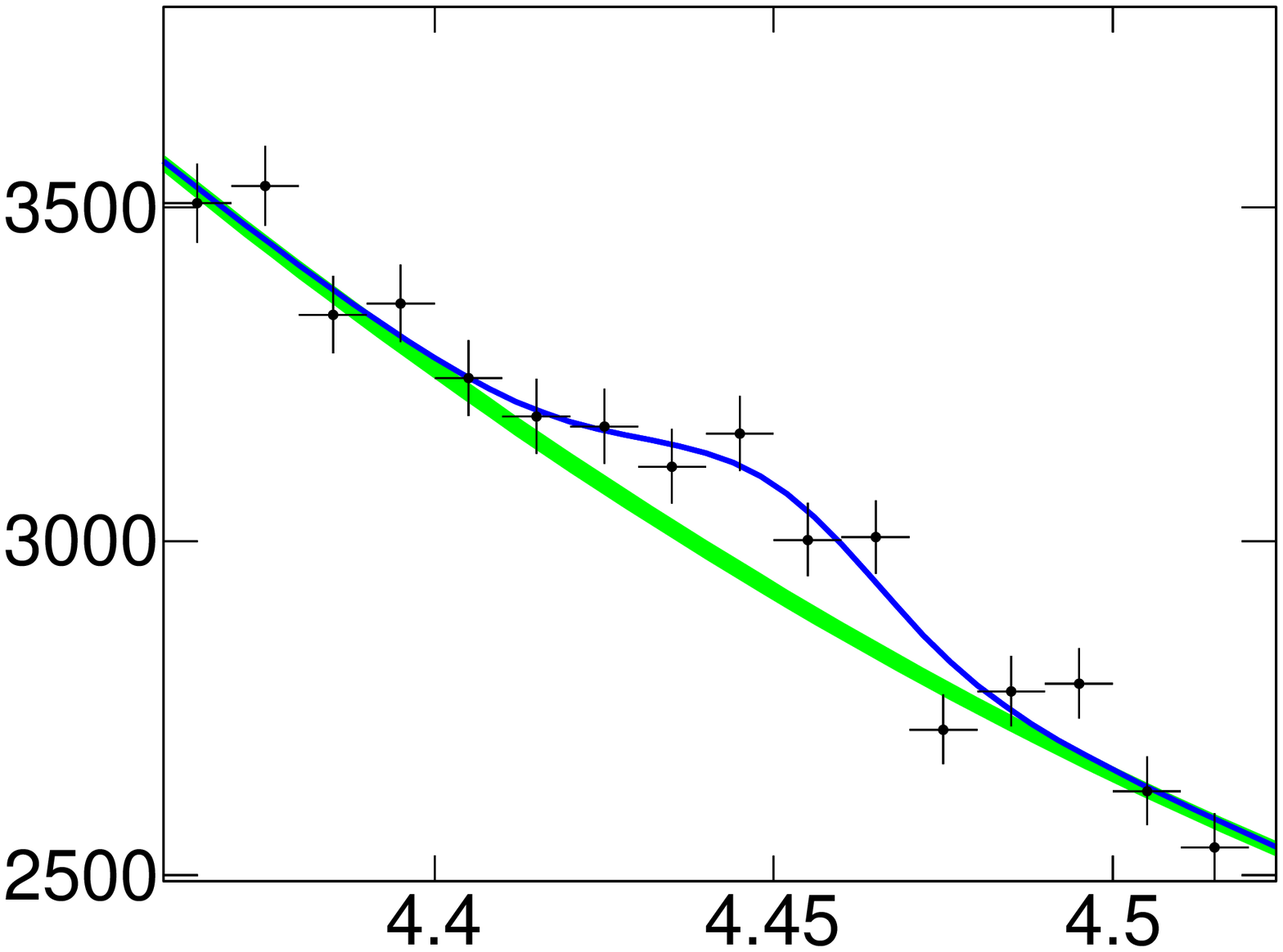}
\vskip1.5cm
   \caption{(color online) Invariant mass distribution of $J/\psi p$ 
``displaced vertex'' candidates
with a superimposed fit that includes an incoherent sum of two Breit-Wigner resonances
 with parameters  set to the values
 reported in Ref.~\cite{lhcb2019}. The ratio of the yields of the two states is allowed to vary
and  background is modeled with a second-order Chebyshev polynomial
(green band). The uncertainty in the background
is represented by the width of the line.   
}
   \label{fig:fixmg}
\end{figure}

With  the baseline background  parametrization,
the fit, shown in Fig.~\ref{fig:fixmg},
gives  a total of  $N=830\pm206$ signal events, The fraction $f=0.61\pm 0.23$
is in agreement with the LHCb value of 0.677.
The probability for background to fluctuate above the observed signal,
 based on the increase of the likelihood with respect
 to the fit with no signal
 $-2\Delta \ln{\cal L}$ of 17.0 for two degrees of freedom~\cite{wilks},
with an assumed positivity constraint,  is 0.000102.
The corresponding statistical significance
 is $S=3.7\sigma$, and the  fit quality is $\chi^2/{\rm ndof}$=36.4/35. This is the baseline fit and measurement.

For a third-order Chebyshev polynomial background,
 the results are  $N=789\pm215$, $f=0.62\pm0.24$,
$S=3.7\sigma$, and $\chi^2/{\rm ndof}$=36.0/34. 
With the ARGUS function of the form
$Argus(m,m_0,s,p) \propto m \cdot [1-(m/m_0)^2]^p\cdot \exp[s\cdot (1-(m/m_0)^2)]$, the fit results are $N=735\pm200$,  $f=0.63\pm0.26$,
$S=3.7\sigma$, and $\chi^2/{\rm ndof}$=36.3/34.
In both cases  the improvement in $\chi^2$ is
less than the penalty~\cite{akaike}  for an additional parameter and thus justifies the choice of the
fit with the second-order Chebyshev polynomial background as the baseline.

We test the  sensitivity to altering single parameters or pairs of parameters with these auxiliary fits.
In these  fits, the fraction $f$ is set to the LHCb value of 0.677.
In all cases the significances differ by no more than 0.1$\sigma$ relative to the baseline fit.
\begin{itemize}
\item
When one width is allowed to vary, with the other set to the LHCb value, the results are
$\Gamma(4440)=32^{+67}_{-27}$~MeV and $\chi^2/{\rm ndof}=36.2/35$, and
$\Gamma(4457)=0^{+35}_{-0}$~MeV and  $\chi^2/{\rm ndof}=36.4/35$.

\item 
A fit allowing  the mass of the lower resonance  to vary and
the other four parameters set to the LHCb values,
gives $M(4440)= 4443^{+8}_{-9}$~MeV  and $\chi^2/{\rm ndof}=36.4/35$.
\item
A fit  in which the mass of the lower resonance is taken as the LHCb central value minus one
standard deviation, obtained as the sum in quadrature of the statistical and systematic uncertainties,
and the  mass  of the higher resonance is similarly shifted  up by 1$\sigma$, 
gives $N=905\pm224$ and $\chi^2/{\rm ndof}=37.6/36$.
\item
A fit in which the lower mass is shifted up by 1$\sigma$
and the higher mass is  shifted down by 1$\sigma$ 
gives $N=805\pm140$  and $\chi^2/{\rm ndof}=36.3/36$.
\end{itemize}

\begin{figure}[htbp]
  \includegraphics[width=0.99\columnwidth]{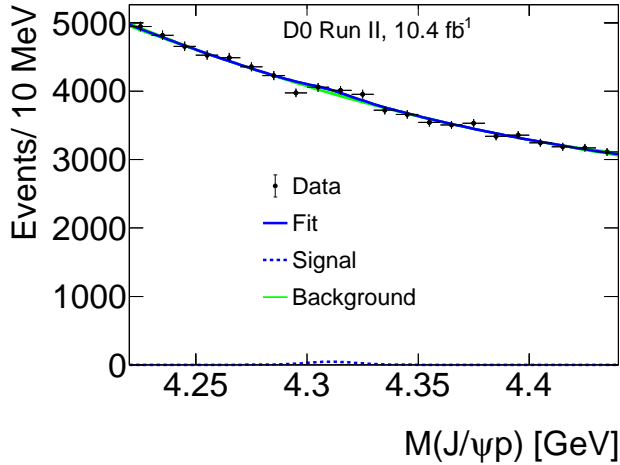}
   \caption{(color online) Invariant mass distribution of $J/\psi p$
 candidates in the vicinity of the $P_c(4312)$.
The fit includes a
 single Breit-Wigner resonance with mass and width
  fixed to the values
of 4311.9~MeV and 9.8~MeV  reported in Ref.~\cite{lhcb2019}
(dotted blue) and a second-order Chebyshev polynomial background
(green $\pm$1$\sigma$ band).
}
   \label{fig:pc4312}
\end{figure}

To search for  the $P_c(4312)$ state  in the ``displaced vertex'' sample,
we perform a fit in the reconstructed mass range 4.22-4.44~GeV, with
 the signal mass and width set to the values of 
4311.9~MeV and 9.8~MeV  reported in Ref.~\cite{lhcb2019}. 
The mass resolution is 9~MeV.
The best fit, shown in Fig.~\ref{fig:pc4312}, with the second-order Chebyshev polynomial background gives $N$ = 151 $\pm$ 186 events.
 The fit quality is $\chi^2/{\rm ndof}=15.5/18$.
The  ratio of the yield of
the $P_c(4312)$ to the sum of $P_c(4440)$ and $P_c(4457)$ at 0.18$\pm$0.22 is
consistent with
the LHCb reported ratio of 
$0.18 \pm 0.06 {\rm \thinspace (stat)} ^{+0.21} _{-0.06} {\rm \thinspace (syst)}$ 
 for the exclusive decay $\Lambda_b^0 \rightarrow J/\psi p K^-$.

\begin{figure}[htbp]
   \includegraphics[width=0.99\columnwidth]{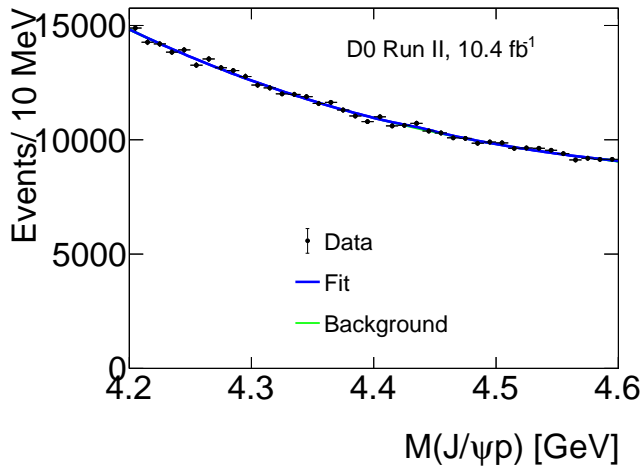}
   \caption{(color online) Invariant mass distribution of ``primary vertex'' $J/\psi p$ candidates.
The fit includes an incoherent sum of two Breit-Wigner resonances with mass and width  parameters  set to the values
 reported in Ref.~\cite{lhcb2019}
and the background modeled with a second-order Chebyshev polynomial
(green $\pm 1\sigma$ band). 
}
   \label{fig:primary}
\end{figure}

For the complementary sample of 451,696 ``primary vertex'' events obtained by reversing
the requirement on the separation  of the $J/\psi p$ vertex from
the primary vertex, the fit assuming
an incoherent sum of the   $P_c(4440)$ and $P_c(4457)$ resonances with fixed
LHCb parameters for the
masses and widths and a free ratio $f$   and a 
second-order polynomial background gives $N=421\pm410$ events 
and $\chi^2/{\rm ndof}=49.2/35$.
The fit is shown in Fig.~\ref{fig:primary}.

The systematic uncertainties in the signal yield for fixed mass and width
are evaluated as follows:

\begin {itemize}

\item Mass resolution

We assign the uncertainty in the signal yields  due to uncertainty in the mass resolution
as half of the difference of the results obtained by changing the resolution 
between  10~MeV and 14~MeV. The fit results for the ``displaced vertex'' sample 
are $N=795\pm 196$ 
and  $N=869\pm222$,  respectively.

\item Background shape

We assign a symmetric uncertainty equal to the
difference of 95 events between the highest and lowest yields  obtained using the three background models.

\item LHCb resonance parameters

We explore the sensitivity of the signal yield
to the parameters of  the two resonances observed in Ref.~\cite{lhcb2019} 
by randomly altering all five parameters using 
the LHCb statistical and systematic uncertainties from Table 1 of Ref.~\cite{lhcb2019}.
We simultaneously vary  the statististical deviations according to  Gaussian distributions in an unlimited range and 
the systematic deviations within $\pm 1\sigma$ assuming uniform distributions.
The choice of the range allowed for the systematic uncertainties is based on the fact the LHCb uncertainties are maximum
 deviations from multiple alternate fits. 
The parameter uncertainties reported in  Ref.~\cite{lhcb2019} assume that the $P_c(4440)$ and $P_c(4457)$ have the same $J^P$ and interfere with an arbitrary phase, thus overestimating the uncertainties for the case of states of different  $J^P$.
The standard deviation of 100 such random alterations is taken as the systematic uncertainty due to 
the LHCb resonance parameters.

\end{itemize}

The systematic uncertainties are shown in Table~\ref{tab:syst}. The total systematic uncertainty on the ``displaced vertex''
event yield, taken as the sum in quadrature, is 128 events.

\begin{table}[htb]
\caption{\label{tab:syst} Systematic uncertainties in  the combined $P_c(4440)$ and $P_c(4457)$
signal yield for ``displaced vertex'' (Fig.~\ref{fig:fixmg}) and
``primary vertex'' (Fig.~\ref{fig:primary}). }

\begin{ruledtabular}
\def\arraystretch{1.3}
\begin{tabular}{lccc}
Source & Displaced vertex  & Primary vertex &\\
\hline
Mass resolution   & $\pm$37  & $\pm$22  & \\
Background shape & $\pm$95 &   $\pm 139$  &\\
 LHCb resonance parameters & $\pm77$ & --&\\
\hline
Total (sum in quadrature) & $\pm 128$   & $\pm 141$ &\\
\end{tabular}
\end{ruledtabular}
\end{table}

To propagate the systematic uncertainties  we  evaluate the $p$-value for the background-only hypothesis
to give   $N$ fitted signal events assuming a Gaussian distribution.
We then convolve the distribution of such $p$-values as a function of $N$ with a normalized Gaussian function with 
a mean of 830 and width $\sigma_N=128$
to get a significance
 of 3.2 $\sigma$.

To obtain the acceptance $A$ of the ``displaced-vertex'' selection for $H_b$
decay events leading to the $P_c$ states,  defined as 
$N_{\rm displaced}/(N_{\rm displaced}+ N_{\rm primary})$,
we use  candidates for the decay  $B^+ \rightarrow J/\psi K^+$
assuming that the distributions of the decay length and its uncertainty for  the $B^+$ decay
 are a good representation for the average $b$ hadron.
All the event selection criteria are the same as for the $P_c$ candidates, except
that the upper limit on $p_T$ of the $J/\psi h^+$ system is removed.
We find  the fitted numbers of $B^+$ decays
$N_{\rm displaced}=46688\pm350$ and  $N_{\rm primary}=12752\pm765$,
and the corresponding acceptance of 0.78.
Relative to the $B^+$ case, the $\Lambda_b$ mean lifetime is 10\% lower
and its mean $p_T$  is about 10\% lower so that the slope of  the transverse decay length,
$L_{\rm xy}$, distribution
is larger than that for $B^+$ by 20\%.
 Assuming that our sample is dominated by $\Lambda_b$, and assuming an
exponential distribution of  $L_{\rm xy}$,  we obtain the acceptance
of $A= 0.73\pm$0.05.

Using the results of the mass fits to the ``displaced-vertex'' and ``primary vertex''  subsamples
we can obtain
acceptance-corrected yields of prompt and nonprompt production
and their ratio. 

The ``displaced-vertex'' signal includes events ($\approx$10\%) with a proton
candidate that originates  from the primary vertex. 
Such events include cases where the proton candidate is ``prompt'' but 
the $J/\psi$ originates from an $H_b$ decay. Thus, the fraction of the
true prompt production of $J/\psi p$ is less than 10\%.
It is accounted for as an additional source
of uncertainty in the calculation of prompt and nonprompt yield but has a negligible impact
 compared to other sources. 

The total yield of the nonprompt
 production is   $N_{\rm nonprompt} = N_{\rm displaced}/A=1136\pm 282$ (stat + syst).
The net number of prompt events is
  $N_{\rm prompt}=N_{\rm primary} - (1-A)\times N_{\rm nonprompt} =114\pm 430 $.
In calculating the uncertainty on the total prompt yield, we add  the statistical
and the systematic uncertainty components  in quadrature.
We obtain  the ratio
 $N_{\rm prompt}/N_{\rm nonprompt}=0.1 \pm 0.4$.
Assuming Gaussian uncertainties and setting the Bayesian prior for negative values of the ratio
to zero,  we obtain an upper limit of 0.9 at the  95\% credibility level.

To test the robustness of the signal in the ``displaced vertex'' data,
 we performed  fits for various alternative selection criteria.
As in the baseline fit, the signal mass and width parameters are set to the LHCb
values, the fraction $f$ is allowed to vary, 
 and the background is modeled by the second-order Chebyshev
polynomial. The signal is present in the entire rapidity range of $(-2,2)$
with yields expected for $b$-hadron decays.
The results for the three regions of $|y|$
of the $J/\psi p$ rapidity, $|y|<0.9$, $0.9<|y|<1.3$, and $|y|>1.3$
are $247\pm111$, $234\pm116$, and $347\pm134$, respectively.
When we increase the upper limit on the  $J/\psi p$  $p_T$ to
14~GeV, the signal yield is increased by 10\% to 915$\pm$243
while the background is increased by 40\%. This is in agreement with
the expectation, due to the difference in the $p_T$ distributions of
the $\Lambda_b^0$ baryons and $B$ mesons. The statistical significance
 of the signal is slightly lowered to  3.3$\sigma$.
For the upper limits of 11 GeV and 13 GeV, the signal yields are  $532 \pm 174$ and  $825 \pm 230$  events, with corresponding statistical significances of $2.7 \sigma$ and $3.3 \sigma$.   
Within statistical fluctuations the variation of these
significances with the $p_T$ limit  conforms to our expectations.

\begin{figure}[htbp]
 \includegraphics[width=0.99\columnwidth]{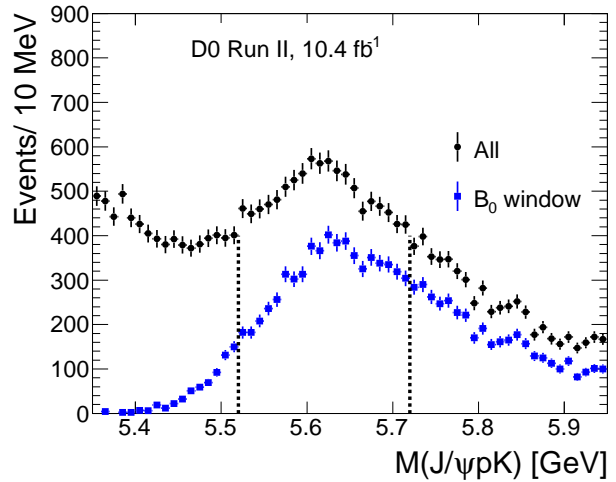}
   \caption{(color online) The  invariant mass distribution of $\Lambda_b^0 \rightarrow J/\psi p K^-$
candidates.
 Also shown (in blue)
is the mass distribution for events in the  $B^0_d$ mass window. The vertical lines
indicate the $\Lambda_b^0$ mass window.
}
   \label{fig:psipkexcl}
\end{figure}

Since the $P_c(4450)$ states were originally observed
 in the $\Lambda_b^0 \rightarrow J/\psi p K^-$ 
channel we should expect to see some indication of them in that exclusive channel.
We have examined a subsidiary sample in which we require that there is an additional negative
 track with a transverse momentum $p_T>0.7$~GeV
 assigned to be a kaon. The addition to the $\chi^2$ of the vertex fit
is required to be less than six. To select events with a displaced vertex, we 
require the  $J/\psi p K^-$  vertex  to be  displaced  in the transverse plane from
the $p\bar{p}$  interaction vertex  by at least 3$\sigma$
and we apply a constraint
on the pointing angle~\cite{angle} of $\alpha<0.06$ radians. The mass distribution for
accepted candidates is shown in Fig.~\ref{fig:psipkexcl}.
The $\Lambda_b^0$ signal region is defined as 5.52--5.72~GeV, corresponding
approximately to $\pm2.5$ standard deviations of our mass resolution.
There is a large peaking background from fully reconstructed 3-body decays of $B$ mesons
 treated as $J/\psi p K^-$,
and a falling background mainly due to multibody $H_b$ decays.

The largest peaking background is due to the decay
 $B^0_d \rightarrow J/\psi K^{\pm} \pi^{\mp}$. 
 There are 8262$\pm$176 $B^0_d$ events contributing to 
the distribution shown in Fig.~\ref{fig:psipkexcl}.
The mass distribution for
events from the $B^0_d$ mass window, 5.15--5.4~GeV, treated as $J/\psi p K^-$,
is shown in blue in Fig.~\ref{fig:psipkexcl}. 
The ATLAS Collaboration presented a similar distribution   
in Fig. 1 of a  conference report ~\cite{atlaspc}.
According to  the ATLAS Monte Carlo  estimates, 
 the ratio of yields of $\Lambda_b  \rightarrow J/\psi p K$ to
$B^0 \rightarrow J/\psi K \pi$ is $\approx$0.2.   .
 Hence, we can estimate the number of  $\Lambda_b \rightarrow J/\psi p K$ decays in our sample
 to be $\approx$1700.
According to  LHCb~\cite{lhcb2019} , the fit fraction for
$\Lambda_b$ decay to the sum of the
two  $P_c$ states is $B=0.0164$.  
 This leads to an expected number of $\Lambda_b \rightarrow P_c K$
events of $\approx$27  in our sample.

\begin{figure}[htbp]
\includegraphics[width=0.99\columnwidth]{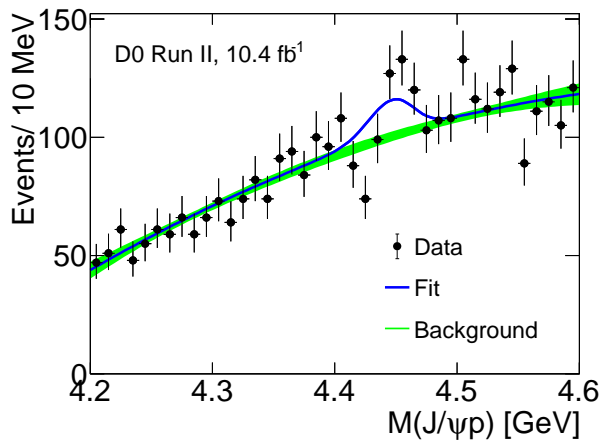}
   \caption{(color online) 
 The  invariant mass distribution of $J/\psi p$ candidates   for
the candidates of the decay  $\Lambda_b^0 \rightarrow J/\psi p K^-$. 
}
   \label{fig:psipexcl}
\end{figure}

A fit to the  sample within  the $\Lambda_b^0$ mass region
with resonance masses and widths and the ratio $f$ set to the LHCb values
 is shown in Fig.~\ref{fig:psipexcl}. 
The fit results are $N=82\pm37$ and $S=2.3\sigma$.
It should be noted that  the falling  background  in Fig.~\ref{fig:psipkexcl}
includes events that have a $J/\psi p$ pair
among the decay products and may also contribute to the $P_c$ signal.
If we require that $M(p K) > 1.9$ GeV   so as to remove events from
$\Lambda_b \rightarrow J/\psi \Lambda^*$, as was imposed by LHCb, the yield
 decreases to $13\pm 14$ events.
The disparity between the estimated number of $\Lambda_b \rightarrow P_c K$
events and the observed number of fitted $P_c$ events in Fig.~\ref{fig:psipexcl}, as well as the reduction in
the number of fitted $P_c$ events when the cut $M(J/\psi p  K^-)>1.9$~GeV is imposed, indicate the
possibility that
 decays $\Lambda_b \rightarrow P_cX$ other than those with $X=K^-$ are present.

In summary, we have studied the inclusive production of the $J/\psi$ meson
associated with a particle assumed to be a proton.
For a subsample of events consistent with coming from decays
of $b$ hadrons, we find an enhancement in the $J/\psi p$ invariant
mass consistent with a sum of  resonances $P_c(4440)$ and $P_c(4457)$
reported in Ref.~\cite{lhcb2019}. 
This is the first confirmatory evidence for these pentaquark states.
The statistical significance of the pentaquark signal with mass and width parameters 
set to the LHCb values is  3.7$\sigma$.
The total significance of the signal obtained with the input parameters set  to the LHCb values
and including the D0 systematic uncertainties and uncertainties in the LHCb input
mass and width  parameters for the $P_c(4440)$ and $P_c(4457)$ is $3.2\sigma$.
The systematic uncertainty due to the LHCb resonance parameters is conservative
since Ref.~\cite{lhcb2019}
  assumes that the $P_c(4440)$ and $P_c(4457)$ are coherent.
The measured ratio $f=N(4440)/(N(4440)+N(4457))$ of $0.61\pm0.23$ is consistent
with the LHCb value.
The study of the semi-exclusive process $\Lambda_b \rightarrow J/\psi p X$
indicates the possibility that decays $\Lambda_b \rightarrow P_c  X$
other than those with  $X=K^-$ exist.

There is no evidence of prompt production of the $P_c(4450)$ states. We find 
$N_{\rm prompt}/N_{\rm nonprompt}=0.1 \pm 0.4$ and 
obtain an upper limit of 0.9 at the  95\% credibility level.

The  ratio of the yield of
the $P_c(4312)$ to the sum of $P_c(4440)$ and $P_c(4457)$ is 0.18$\pm$0.22 which
is consistent with
the value measured by  LHCb. 

%
We thank the staffs at Fermilab and collaborating institutions,
and acknowledge support from the
DOE and NSF (USA);
CEA and CNRS/IN2P3 (France);
MON, NRC KI and RFBR (Russia);
CNPq, FAPERJ, FAPESP and FUNDUNESP (Brazil);
DAE and DST (India);
Colciencias (Colombia);
CONACyT (Mexico);
NRF (Korea);
FOM (The Netherlands);
STFC and the Royal Society (United Kingdom);
MSMT and GACR (Czech Republic);
BMBF and DFG (Germany);
SFI (Ireland);
The Swedish Research Council (Sweden);
and
CAS and CNSF (China).


\begin{thebibliography}{99}
\bibitem{eta}
$\eta=-\ln[\tan(\theta/2)]\ $ is the pseudorapidity and $\theta$ is
the polar angle between the track momentum and the proton beam direction.
\bibitem{ip}
 The impact parameter IP is defined as the distance of closest approach of the track to the $p\bar{p}$
 collision point projected onto the plane transverse to the $p\bar{p}$ beams.
\bibitem{angle}
The pointing angle is defined as the angle in the $x-y$ plane
between a particle's
momentum vector and the vector from the primary vertex
to the particle's decay vertex.
\end{thebibliography}
\end{document}